\documentstyle[epsf,prb,aps]{revtex}
\begin{document}
\draft

\title{Fingering Instability of Dislocations and Related Line Defects}
\author{Ming Li$^{*\dagger}$, Brian B. Smith$^*$, and Robin L. B. Selinger$^*$}
\address{$^*$Physics Department, Catholic University of America,
Washington, DC 20064}
\address{$^\dagger$ Center for Nonlinear Science, Los Alamos National
Laboratory, Los Alamos, New Mexico 87545}
\date{\today}
\maketitle
\begin{abstract}

We identify a fundamental morphological instability of mobile
dislocations in crystals and related line defects. A positive gradient
in the local driving force along the direction of defect motion
destabilizes long-wavelength vibrational modes, producing a
``fingering'' pattern. The minimum unstable wavelength scales as the
inverse square root of the force gradient.  We demonstrate the
instability's onset in simulations of a screw dislocation in Al (via
molecular dynamics) and of a vortex in a 3-d XY ``rotator'' model.

\end{abstract}

\def\del{\partial}

Modeling the motion of dislocations in crystals is mathematically
challenging due to their long-range interaction through the elastic
strain field~\cite{hirth}, and computational approaches include
atomistic molecular dynamics~\cite{zhou,bulatov} and mesoscale
models~\cite{kubin,schwarz}. In some contexts the complexity of the
problem can be reduced by approximating the motion of an individual
dislocation as that of a damped elastic string~\cite{granato}, and
such models can be used to describe the motion of other types of line
defects as well, e.g. magnetic flux lines in type-II
superconductors. Recent research using this approach has focused on
locked vs. running behavior~\cite{cattuto} of a line defect in a
periodic potential due to the underlying crystal lattice. A morphological
instability associated with a periodic driving force at a frequency
resonant with the periodic potential has also been studied~\cite{kolomeisky}.

In the present Letter we identify a fingering instability in the glide
of a mobile line defect under a spatially varying driving force that
{\it increases} along the direction of motion.  This work was
motivated by unpublished molecular dynamics studies of screw
dislocation pair annihilation in pure Cu performed by Swaminarayan and
LeSar \cite{lesar}.  The simulations appeared to show that initially
straight parallel screw dislocations of opposite Burgers vector
spontaneously developed a wavy profile prior to pair annihilation,
something like the well known Crow instability of parallel vortices in
fluid dynamics \cite{crow}. We were motivated by this result to
consider whether the instability is driven by the attractive force
gradient between the dislocations, and indeed whether a single
dislocation could be unstable under a positive gradient in the driving
stress. Since such stress gradients arise from common
sources--e.g. external loading or dislocation interactions with a free
surface or other microstructural features--such an instability might
be active during metal deformation.

Consider the stability of an initially straight dislocation under an
attractive force gradient, approximating its response as that of a
damped elastic string. (The approximation is useful only to model the
onset of any instability, because once the defect line deviates
strongly from its initial state then nonlocal interactions can no
longer be neglected.) The string's displacement obeys the familiar
wave equation with linear damping and a spatially varying driving
force:
\begin{equation}
{{\del^2 \Psi}\over{\del t^2}}=c^2 {{\del^2 \Psi}\over{\del z^2}} -
{{\gamma}\over{m}}  {{\del \Psi}\over{\del t}} + {F(z,t) \over m}.
\end{equation}
Here $\Psi(z,t)$ is the displacement of the string in the $x$
direction, $c$ is the wave speed of an oscillation on the string,
$\gamma$ is a linear damping coefficient, $m$ is the mass per unit
length, and $F(z,t)$ is the driving force per unit length along the
string. We take periodic boundary conditions with $z$
in the interval $(0,L)$.  For our present purposes we do not
include an underlying periodic potential; this is
appropriate for modeling dislocation motion in FCC metals where the
Peierls stress is small and unpinned dislocations are mobile even at
low temperature.

We consider a driving force with a value that increases linearly in the
direction the string is moving, as if the string were moving down a hill
that gets continuously steeper. The force on the string at any
point along its length may then be written:
\begin{equation}
F(z,t)= F_0 + F_1 \Psi(z,t).
\label{ForceExpansion}
\end{equation}
Any portion of the string that gets slightly ahead of the rest feels a
stronger driving force. More generally Eqn.~(\ref{ForceExpansion})
represents the first two terms in an expansion, so that $F_1$ is
defined as the local gradient of a spatially dependent driving force.
If $F_1 > 0$ the force gradient is destabilizing and will cause a
sinusoidal perturbation of sufficiently long wavelength to grow rather
than die away, while small wavelength perturbations are suppressed by
line tension, as shown below.

To linearize the equation we switch to a (non-inertial) moving
reference frame. Consider perturbations $\tilde \Psi(z,t)$ from a
stable uniform running solution $\Psi_0(t)$,
\begin{equation}
\Psi(z,t)=\Psi_0 (t) + \tilde \Psi(z,t)
\end{equation}
where
\begin{equation}
{{\del^2 \Psi_0}\over{\del t^2}}=
-{{\gamma}\over{m}}  {{\del \Psi_0}\over{\del t}} + {F_0 \over m} 
+ {{F_1} \over m}\Psi_0.
\end{equation}
This procedure gives rise to the linearized equation of motion for the
perturbation:
\begin{equation}
{{\del^2 {\tilde \Psi}}\over{\del t^2}}=
   c^2 {{\del^2 {\tilde \Psi}}\over{\del z^2}} 
-{{\gamma}\over{m}}  {{\del {\tilde \Psi}}\over{\del t}} + {F_1 \over
   m} {\tilde \Psi}.
\label{Linearized}
\end{equation}
Taking a solution of the form ${\tilde \Psi}
= A {\rm e}^{i(k z-\omega t)},$ we find the
dispersion relation
\begin{equation}
\omega(k)={ - {{i \gamma} \over {2m}} \pm {1 \over 2}\sqrt{{{-\gamma^2}
\over {m^2}} - 4 \left({{F_1}\over{m}}-c^2 k^2 \right)}}. 
\end{equation}
We identify three regions of $k$.

(1) Underdamped region: short wavelength ($k>k_d$, or $\lambda<
\lambda_d$) perturbations are underdamped, with
\begin{equation}
k_d = \sqrt{{ {\gamma^2} \over {4 c^2 m^2}}
+ {{F_1} \over {m c^2}}}.
\end{equation}
We identify a characteristic length $\lambda_d = {{2 \pi}/{k_d}}$ as the
wavelength associated with critical damping; modes with smaller
wavelength are underdamped. 

(2) Overdamped region: for intermediate wavelength (${k^*} < k < {k_d}$,
or $\lambda_d < \lambda < \lambda^*$)
perturbations are overdamped. The cut-off value $k^*$ is
\begin{equation}
k^*= \sqrt{{{F_1}\over{m c^2}}}.
\end{equation}
We identify the characteristic length $\lambda^*$
\begin{equation}
\lambda^*={{2 \pi}\over k^*}=2 \pi \sqrt{{{m c^2}\over{F_1}}}.
\end{equation}
\noindent Note that $\lambda^*$ decreases as the stress gradient
$F_1$ increases, and that it is {\it independent} of the damping
constant $\gamma$.

(3) Unstable region: long wavelength modes ($k < k^*$, or $\lambda >
\lambda^*$) are unstable, and will grow, leading to the formation of a
``fingering'' pattern. The fastest growing mode (which has largest
positive Im $\omega$) is always that with the largest unstable wavelength.

To evaluate these characteristic lengths for a real material, we
consider the case of a mobile screw dislocation in pure Al.  The
damping coefficient $\gamma$ has been estimated at about $10^{-4}$
Pa-sec \cite{gotfromkubin}, but in molecular dynamics simulation at
temperatures close to zero we have found it to be about an order of
magnitude smaller \cite{inprep}. The wave speed is $c=29$ \AA/psec and
the effective mass depends logarithmically on system size but is in
the range of 5 amu/\AA~for a system of about $300 \times 300$ \AA$^2$
in cross section, again as found in a molecular dynamics simulation
\cite{inprep}. In the absence of a stress gradient, the wavelength
associated with critical damping at low temperature is $\lambda_d=$0.2
to 0.3 $\mu{\rm m}$; at higher temperatures where the damping is
stronger, $\lambda_d$ may be as small as a few hundred \AA. The
minimum unstable wavelength $\lambda^*$ depends on the local force
gradient. For alumimum we estimate the energy/length of a screw
dislocation as roughly 1 eV/\AA, and the driving force is $\sigma
\cdot b$, where $\sigma$ is the driving stress and $b$ is the Burgers
vector=2.86 \AA. Then the characteristic unstable wavelength can be
estimated roughly as ${\lambda^*}=5 \times 10^{-4} {\rm m}
{(\sigma/l)}^{-1/2}$ where the stress gradient $\sigma/l$ is in units
of GPa/m. Thus for a screw dislocation in Al, we estimate that a
stress gradient of 1 GPa/cm gives a minimum unstable wavelength of
order ${\lambda^*}= 50 \mu{\rm m}$, and a stress gradient of 100
GPa/cm gives ${\lambda^*}= 5 \mu{\rm m}$.

If $\lambda^*<L$, where $L$ is the length of the defect line, then all
vibrational modes with wavelength $\lambda > \lambda^*$ will be
unstable.  While the lowest unstable mode grows the fastest, in the
early stages of growth all unstable modes are visible and give rise to
a disordered fingering structure. If we follow the evolution of the
linearized equation of motion in Eqn. (\ref{Linearized}) beyond the
early stages of growth, we observe a coarsening process in which
neighboring peaks in the structure merge until only a single sinusoid
profile remains.  In Fig.~1 we show the results of a numerical
integration of Eqn.~(\ref{Linearized}), showing the type of patterns
that arise, where the initial condition was a straight line with a
small amplitude random perturbation.

For dislocations in metals, the linearized equation is only valid to
describe the very earliest stage of motion, so we would not
necessarily expect to see fingering behavior like that shown in
Fig.~1. However the initial growth of the instability should be
described by the analysis above. To test this hypothesis, we performed
a molecular dynamics simulation of the motion of a screw dislocation
in a single crystal of aluminum using an Embedded Atom Method
potential provided by Diana Farkas et al \cite{farkas}.  To see the
instability in an atomic scale simulation is not an easy task, because
to get a minimum unstable wavelength as small as 500 \AA~in a bulk
solid, we would need an enormous stress gradient in the range of
$10^8$ GPa/m. Luckily, there is a much easier way: we make use of the
force gradient produced by the attraction between a dislocation line
and a nearby free surface.

The force/length on a screw dislocation at a distance $R$ from a
parallel free surface scales as $1/R$, so the gradient in the driving
force diverges as $1/R^2$. When the dislocation is far enough from the
surface, the local force gradient is small, the minimum unstable
wavelength is longer than the dislocation line, and the dislocation is
stable.  At a critical distance from the wall ${R^*}$ (which depends
on dislocation length $L$) the dislocation's first vibrational mode
becomes unstable and fingers toward the free surface.  As the
dislocation continues to move toward the free surface, the force
gradient gets stronger and higher order vibrational modes go unstable
one by one in a cascade of instabilities.

Our simulated system has periodic boundary conditions only along the
z-axis and thus has four free surfaces; the dislocation is initially
parallel to the z-axis which lies in the [110] direction. The system
size is $ 45 \times 45 \times 183 = 370,575$ atoms, or $111 \times 105
\times 524$ {\AA}$^3$, and the temperature is close to $0^o$ K. The
dislocation is placed a distance of 26.5 \AA~from one free surface and
is given a very slight sinusoidal perturbation with wavelength
$\lambda=L=524$\AA. We estimate the minimum unstable wavelength as:
\begin{equation}
\lambda^*= 2 \pi \left[ { \left( {\rm log}({{2R}\over{a_o}})+{\rm
log}({{2{D}-2R}\over{a_o}})-{\rm log}({{2{D}}\over{a_o}}) \right)}
\over { R^{-2} +(D-R)^{-2}} \right]^{1/2}
\label{Lambdastar}
\end{equation}
\noindent where the numerator is the energy/length of the dislocation
($mc^2$) and the denominator is the gradient of the driving
force/length\cite{hirth}, using image terms from free surfaces ahead
and behind, and canceling the energy prefactor. Here $D=111$\AA~is the
length of the crystal in the direction of dislocation motion, and
$a_o=b$ is the dislocation core size.  For a dislocation at position
$R=26.5 \AA,$ Eqn. (\ref{Lambdastar}) predicts $\lambda^*=258$ \AA. Since
$\lambda^*$ is about half the size of the defect length $L=524$ \AA,
we expect both the first and second vibrational modes to be unstable
and grow.

The dislocation is shown in Fig.~2, where we display only atoms whose
local potential energy differs from that of the bulk. The initial
state is shown in gray, the later state in black.  The dislocation
splits into two partial dislocations separated by a ribbon of stacking
fault. In each of the partials, the initial sinusoidal perturbation
grows along with the second vibrational mode, and each of the partials
fingers toward the wall. The apparent minimum unstable wavelength is
roughly $\lambda^*=250$ \AA, in agreement with our prediction.  We
have also run a simulation of the same set-up performed in a crystal
of height $150$ \AA; in this case the initial perturbation does not
grow, as its wavelength $\lambda=L<\lambda^*$, and the instability is
suppressed by finite size.

Our simulation of a dislocation moving toward a free
surface is closely related to Swaminarayan's \cite{lesar}
study of dislocation pair annihilation. In our case, the
dislocation is attracted to an ``image'' dislocation \cite{hirth} of
opposite sign located in a symmetric position across the free
surface. In order to see the longer time evolution of the instability,
we need a larger system, so we have 
performed simulations of vortex motion in a three-dimensional XY
``rotator'' model of a simple magnet \cite{romano}. This system is
very efficient computationally, and we have run simulations of up to
$4 \times 10^6$ spins on a DEC Alpha workstation. In this model,
each spin on a cubic lattice has a moment of inertia $I=1$ and one
rotational degree of freedom with angular position $\theta_i$ and
angular velocity $\omega_i$. The spins have nearest-neighbor
interactions with an XY Hamiltonian, an applied magnetic field
$B$, plus a rotational kinetic energy term:
\begin{equation}
H=-\sum_{<i,j>}\cos({\theta_i}-{\theta_j}) + \sum_{i} 
-B \cos(\theta_i) + {1 \over 2} I
{\omega_i}^2. 
\label{Hamiltonian}
\end{equation}
\noindent We derive equations of motion from
Eqn. (\ref{Hamiltonian}) under constant energy. A straight vortex line
has properties analogous to those of a screw dislocation in a crystal,
particularly the effective mass/length, mobility behavior, Peierls
barrier, and vortex-vortex interactions \cite{inprep}.

We use periodic boundary conditions along the z direction only with
four free surfaces, and size $100 \times 100 \times 400$ spins.  In
the initial state a vortex line lies parallel to the z axis at an
initial temperature ${k_B}T=0.25$, in the ordered phase. Since the
simulation is run at constant energy, $k_BT$ fluctuates. We
apply a small exteral field B=-0.003 which drives the vortex in the +x
direction.

In Fig.~3 we show the time evolution of the vortex,
starting at x=40 and moving toward the free surface
at x=100.  Using Eqn.(\ref{Lambdastar}) with $a_o=1$ and
$D=100$, we predict that the first vibrational mode should become
unstable when the vortex reaches a distance $R=38$ lattice sites from
the free surface.  Since this system is at finite temperature,
the vortex line roughens initially due to
thermal fluctuations (like those studied by Chzran and Daw
\cite{chrzan} in dislocations.) The first vibrational mode starts to
grow at about the predicted location, around x=60 ($R=40$ units from
the surface.)  Around $x=65,$ the second mode becomes prominent (with
two ``peaks'' visible) and grows until the vortex annihilates at the
free surface. The second mode appears earlier than predicted by
Eqn. (\ref{Lambdastar}); perhaps nonlinear effects
play an important role. In future work we
plan to simulate a larger system in which more steps in the cascade of
instabilities may be seen.

The physical signifigance of this instability is an open question.
Most important, we have learned how finite size effects play a major
role in simulations of line defects in general; and that the {\it
largest} characteristic length is defined by the {\it smallest} force
gradient present. Beyond that we speculate that the characteristic
length scale $\lambda^*$ might play a role in the formation of
dislocation patterns during work hardening. The instability could also
control pair annihilation of threading dislocations in thin films,
defining an effective ``kill radius'' that depends on film thickness
\cite{inprep}.  Confirming any of these mechanisms via experiment may
be difficult. The motion of an individual dislocation in a thin metal
sample can be observed via TEM, but it is not clear whether the
necessary time resolution and length scales are technically
feasible. It would be easier to observe the motion of a defect line in
a liquid crystal driven by a gradient in an applied field, since
defect motion can be viewed with an optical microscope and polarizers.

A closely related instability has previously been identified by Z. Suo
\cite{suo} who noted that a dislocation moving via climb can be
unstable in the presence of a gradient in the density of
vacancies. However since climb is a diffusive process, the dynamics of
that instability mechanism occur on a far slower time scale where
inertial effects are not important. However both mechanisms should
play an important role in dislocation pattern formation in metals.

We thank Lyle Levine, Robb Thompson, and Shujia Zhou for helpful
discussions. This work was supported by NSF Grant No. DMR-59702234.

\begin{figure}
\centering\leavevmode\epsfxsize=5in\epsfbox{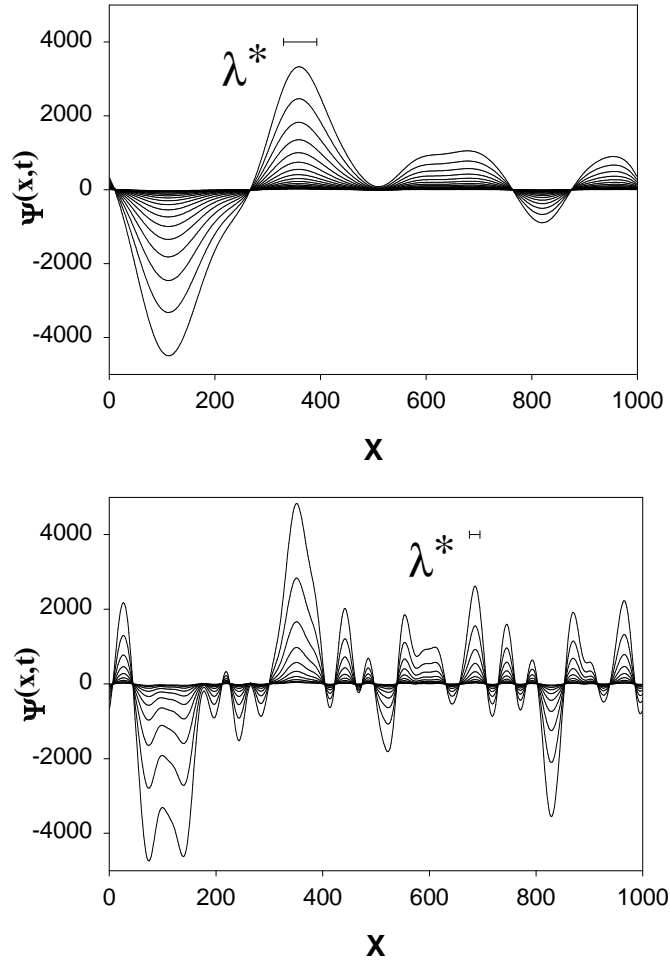}

\vskip2pc
\caption{Evolution of a line defect from numerical
integration of Eqn. (10), $\gamma=0.1, c=1, m=1$, and $L=1000.$
(a) With a weak force gradient $F_1=0.01$, $\lambda^*=63.$
(b) With stronger force gradient $F_1=0.1$,
$\lambda^*=20$. The pattern coarsens via tip combination.}

\label{fig1}
\end{figure}

\begin{figure}
\centering\leavevmode\epsfxsize=5in\epsfbox{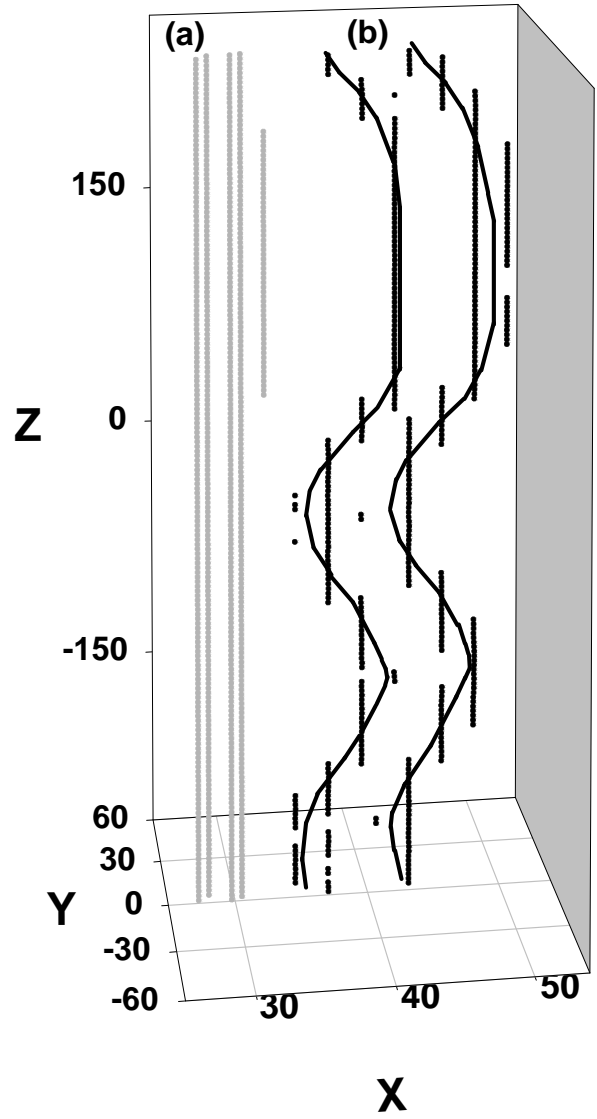}
\vskip2pc
\caption{Molecular dynamics simulation of screw dislocation motion
in Al. (a) Initial configuration. (b) Later configuration; 
dislocation has split into partials which ``finger'' toward the
free surface. Curves are drawn to guide the eye.}
\label{fig2}
\end{figure}

\begin{figure}
\centering\leavevmode\epsfxsize=5in\epsfbox{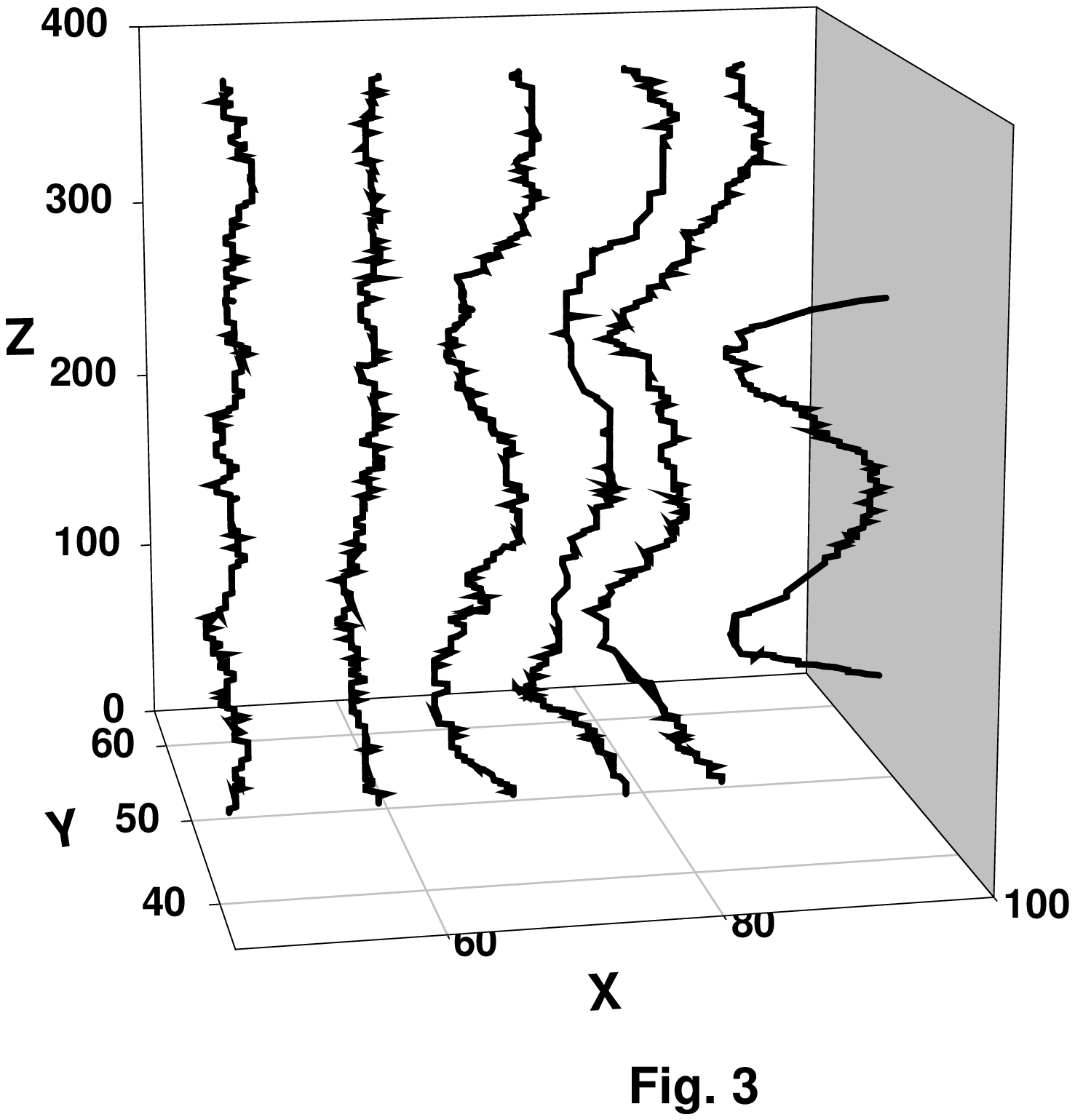}
\vskip2pc
\caption{Time evolution of vortex motion in a 3-d XY ``rotator''
model simulation. As the vortex line
approaches the free surface the first and
then second vibrational modes become unstable.}
\label{fig3}
\end{figure}


\begin{references}

\bibitem{hirth} {For a general introduction to dislocations and their
interactions, see J. P. Hirth and J. Lothe, Theory of Dislocations
(Krieger Publishing, Florida, 1992).}

\bibitem{zhou} {S. J. Zhou, D. M. Beazley, P. S. Lomdahl, and B. L.Holian,
{\it Phys. Rev. Lett.} {\bf 78}, 479 (1997); S. J. Zhou, * D. L. Preston,
P. S. Lomdahl, D. M. Beazley, {\it Science} {\bf 279}, 1525 (1998).}

\bibitem{bulatov} {V. Bulatov, F. Abraham, L. Kubin, B. Devincre,
S. Yip, {\it Nature} {\bf 391}, 669 (1998).}

\bibitem{kubin} {L. P. Kubin et al., Solid State Phenom. 23, 455 (1992).}

\bibitem{schwarz}{K. W. Schwarz, {\it Phys. Rev. Lett.} {\bf 78}, 4785
(1997); K. W. Schwarz and F. K. LeGoues, {\it Phys. Rev. Lett.} {\bf
79}, 1877 (1997).}

\bibitem{granato} {A. Granato and K. Lucke, {\it J. Appl. Phys.} {\bf 27}, 583
(1956); {\it J. Appl. Phys.} {\bf 27}, 789 (1956).}

\bibitem{cattuto} {C. Cattuto and F. Marchesoni,
{\it Phys. Rev. Lett.} {\bf 79}, 5070 (1997).}

\bibitem{kolomeisky}{E. B. Kolomeisky, T. Curcic,and J. P. Straley,
{\it Phys. Rev. Lett.} {\bf 75}, 1775  (1995).}

\bibitem{lesar} S. Swaminarayan and R. LeSar, presented at the
Combined Workshops on Fracture, Friction, and Deformation, held at the
Center for Nonlinear Studies, Los Alamos National Laboratory, Los
Alamos, New Mexico, April 19-24, 1996.

\bibitem{crow}{S. C. Crow, AIAA {\bf 8}, 2172 (1970).}


\bibitem{gotfromkubin}{L. Kubin, private communication.}

\bibitem{inprep}{M. Li, Jianling Xu, Brian Smith, and R. L. B. Selinger, in preparation.}

\bibitem{farkas}Y. Mishin, M. Mehl, D. Farkas and D.
Papaconstantopoulos,  to appear, Phys Rev. {\bf B}.


\bibitem{romano}{For a study of a related two-dimensional system, see 
S. Romano, Phys. Rev. B {\bf 42}, 8647 (1990).}

\bibitem{chrzan}{ D. C. Chrzan and M. S. Daw, Phys. Rev. {\bf B} 55,
798 (1997).} 


\bibitem{suo}{Z. Suo, Acta Met.  {\bf 42}, 3581 (1994).}




\end{references}
\end{document}